\documentclass[12pt,preprint]{aastex}
\newcommand{\etal}{et al.}
\begin{document}
\title{The Fifth Data Release of the Sloan Digital Sky Survey}

\author{
Jennifer K. Adelman-McCarthy\altaffilmark{\ref{Fermilab}},
Marcel A. Ag\"ueros\altaffilmark{\ref{Washington}},
Sahar S. Allam\altaffilmark{\ref{Fermilab},\ref{Wyoming}},
Kurt S. J. Anderson\altaffilmark{\ref{APO}},
Scott F. Anderson\altaffilmark{\ref{Washington}},
James Annis\altaffilmark{\ref{Fermilab}},
Neta A. Bahcall\altaffilmark{\ref{Princeton}},
Coryn A. L. Bailer-Jones\altaffilmark{\ref{MPIA}},
Ivan K. Baldry\altaffilmark{\ref{LJMU},\ref{JHU}},
J. C. Barentine\altaffilmark{\ref{APO}},
Timothy C. Beers\altaffilmark{\ref{MSUJINA}},
V. Belokurov\altaffilmark{\ref{Cambridge}},
Andreas Berlind\altaffilmark{\ref{NYU}},
Mariangela Bernardi\altaffilmark{\ref{Penn}}, 
Michael R. Blanton\altaffilmark{\ref{NYU}},
John J. Bochanski\altaffilmark{\ref{Washington}},
William N. Boroski\altaffilmark{\ref{Fermilab}},
D. M. Bramich\altaffilmark{\ref{Cambridge}},
Howard J. Brewington\altaffilmark{\ref{APO}},
Jarle Brinchmann\altaffilmark{\ref{Porto}},
J. Brinkmann\altaffilmark{\ref{APO}},
Robert J. Brunner\altaffilmark{\ref{Illinois}},
Tam\'as Budav\'ari\altaffilmark{\ref{JHU}},
Larry N. Carey\altaffilmark{\ref{Washington}},
Samuel Carliles\altaffilmark{\ref{JHU}},
Michael A. Carr\altaffilmark{\ref{Princeton}},
Francisco J. Castander\altaffilmark{\ref{Barcelona}},
A. J. Connolly\altaffilmark{\ref{Pitt}},
R. J. Cool,\altaffilmark{\ref{Arizona}},
Carlos E. Cunha,\altaffilmark{\ref{Chicago},\ref{CfCP}},
Istv\'an Csabai\altaffilmark{\ref{Eotvos},\ref{JHU}},
Julianne J. Dalcanton\altaffilmark{\ref{Washington}},
Mamoru Doi\altaffilmark{\ref{IoaUT}},
Daniel J. Eisenstein\altaffilmark{\ref{Arizona}},
Michael L. Evans\altaffilmark{\ref{Washington}},
N. W. Evans\altaffilmark{\ref{Cambridge}},
Xiaohui Fan\altaffilmark{\ref{Arizona}},
Douglas P. Finkbeiner\altaffilmark{\ref{Princeton}},
Scott D. Friedman\altaffilmark{\ref{STScI}},
Joshua A. Frieman\altaffilmark{\ref{Fermilab},\ref{Chicago},\ref{CfCP}},
Masataka Fukugita\altaffilmark{\ref{ICRRUT}},
Bruce Gillespie\altaffilmark{\ref{APO}},
G. Gilmore\altaffilmark{\ref{Cambridge}},
Karl Glazebrook\altaffilmark{\ref{JHU}},
Jim Gray\altaffilmark{\ref{Microsoft}},
Eva K. Grebel\altaffilmark{\ref{Basel}},
James E. Gunn\altaffilmark{\ref{Princeton}},
Ernst de Haas\altaffilmark{\ref{Princeton}},
Patrick B. Hall\altaffilmark{\ref{York}},
Michael Harvanek\altaffilmark{\ref{APO}},
Suzanne L. Hawley\altaffilmark{\ref{Washington}},
Jeffrey Hayes\altaffilmark{\ref{Catholic}},
Timothy M. Heckman\altaffilmark{\ref{JHU}},
John S. Hendry\altaffilmark{\ref{Fermilab}},
Gregory S. Hennessy\altaffilmark{\ref{USNO}},
Robert B. Hindsley\altaffilmark{\ref{NRL}},
Christopher M. Hirata\altaffilmark{\ref{IAS}},
Craig J. Hogan\altaffilmark{\ref{Washington}},
David W. Hogg\altaffilmark{\ref{NYU}},
Jon A. Holtzman\altaffilmark{\ref{NMSU}},
Shin-ichi Ichikawa\altaffilmark{\ref{NAOJ}},
Takashi Ichikawa\altaffilmark{\ref{Tohoku}},
\v{Z}eljko Ivezi\'{c}\altaffilmark{\ref{Washington}},
Sebastian Jester\altaffilmark{\ref{Southampton}},
David E. Johnston\altaffilmark{\ref{JPL},\ref{Caltech}},
Anders M. Jorgensen\altaffilmark{\ref{LANL}},
Mario Juri\'{c}\altaffilmark{\ref{Princeton},\ref{IAS}},
Guinevere Kauffmann\altaffilmark{\ref{MPA}},
Stephen M. Kent\altaffilmark{\ref{Fermilab}},
S. J. Kleinman\altaffilmark{\ref{Subaru}},
G. R. Knapp\altaffilmark{\ref{Princeton}},
Alexei Yu. Kniazev\altaffilmark{\ref{MPIA}},
Richard G. Kron\altaffilmark{\ref{Chicago},\ref{Fermilab}},
Jurek Krzesinski\altaffilmark{\ref{APO},\ref{MSO}},
Nikolay Kuropatkin\altaffilmark{\ref{Fermilab}},
Donald Q. Lamb\altaffilmark{\ref{Chicago},\ref{EFI}},
Hubert Lampeitl\altaffilmark{\ref{STScI}},
Brian C. Lee\altaffilmark{\ref{LBL},\ref{Gatan}},
R. French Leger\altaffilmark{\ref{Fermilab}},
Marcos Lima,\altaffilmark{\ref{ChicagoPhys},\ref{CfCP}},
Huan Lin\altaffilmark{\ref{Fermilab}},
Daniel C. Long\altaffilmark{\ref{APO}},
Jon Loveday\altaffilmark{\ref{Sussex}},
Robert H. Lupton\altaffilmark{\ref{Princeton}},
Rachel Mandelbaum\altaffilmark{\ref{IAS}},
Bruce Margon\altaffilmark{\ref{SantaCruz}},
David Mart\'{\i}nez-Delgado\altaffilmark{\ref{IAC}},
Takahiko Matsubara\altaffilmark{\ref{Nagoya}},
Peregrine M. McGehee\altaffilmark{\ref{LANL2}},
Timothy A. McKay\altaffilmark{\ref{Michigan}},
Avery Meiksin\altaffilmark{\ref{Edinburgh}},
Jeffrey A. Munn\altaffilmark{\ref{NOFS}},
Reiko Nakajima\altaffilmark{\ref{Penn}},
Thomas Nash\altaffilmark{\ref{Fermilab}},
Eric H. Neilsen, Jr.\altaffilmark{\ref{Fermilab}},
Heidi Jo Newberg\altaffilmark{\ref{RPI}},
Robert C. Nichol\altaffilmark{\ref{Portsmouth}},
Maria Nieto-Santisteban\altaffilmark{\ref{JHU}},
Atsuko Nitta\altaffilmark{\ref{Gemini}},
Hiroaki Oyaizu,\altaffilmark{\ref{Chicago},\ref{CfCP}},
Sadanori Okamura\altaffilmark{\ref{DoAUT}},
Jeremiah P. Ostriker\altaffilmark{\ref{Princeton}},
Nikhil Padmanabhan\altaffilmark{\ref{Princetonphys},\ref{LBL}},
Changbom Park\altaffilmark{\ref{KIAS}},
John Peoples Jr.\altaffilmark{\ref{Fermilab}},
Jeffrey R. Pier\altaffilmark{\ref{NOFS}},
Adrian C. Pope\altaffilmark{\ref{JHU}},
Dimitri Pourbaix\altaffilmark{\ref{Princeton},\ref{Bruxelles}},
Thomas R. Quinn\altaffilmark{\ref{Washington}},
M. Jordan Raddick\altaffilmark{\ref{JHU}},
Paola Re Fiorentin\altaffilmark{\ref{MPIA}},
Gordon T. Richards\altaffilmark{\ref{JHU},\ref{Drexel}},
Michael W. Richmond\altaffilmark{\ref{RIT}},
Hans-Walter Rix\altaffilmark{\ref{MPIA}},
Constance M. Rockosi\altaffilmark{\ref{Lick}},
David J. Schlegel\altaffilmark{\ref{LBL}},
Donald P. Schneider\altaffilmark{\ref{PSU}},
Ryan Scranton\altaffilmark{\ref{Pitt}},
Uro\v{s} Seljak\altaffilmark{\ref{Princetonphys},\ref{Princeton}},
Erin Sheldon\altaffilmark{\ref{Chicago},\ref{CfCP}},
Kazu Shimasaku\altaffilmark{\ref{DoAUT}},
Nicole M. Silvestri\altaffilmark{\ref{Washington}},
J. Allyn Smith\altaffilmark{\ref{LANL},\ref{APeay}},
Vernesa Smol\v{c}i\'{c}\altaffilmark{\ref{MPIA}},
Stephanie A. Snedden\altaffilmark{\ref{APO}},
Albert Stebbins\altaffilmark{\ref{Fermilab}}
Chris Stoughton\altaffilmark{\ref{Fermilab}},
Michael A. Strauss\altaffilmark{\ref{Princeton}},
Mark SubbaRao\altaffilmark{\ref{Chicago},\ref{Adler}},
Yasushi Suto\altaffilmark{\ref{TokyoPhys}},
Alexander S. Szalay\altaffilmark{\ref{JHU}},
Istv\'an Szapudi\altaffilmark{\ref{Hawaii}},
Paula Szkody\altaffilmark{\ref{Washington}},
Max Tegmark\altaffilmark{\ref{MIT}},
Aniruddha R. Thakar\altaffilmark{\ref{JHU}},
Christy A. Tremonti\altaffilmark{\ref{Arizona}},
Douglas L. Tucker\altaffilmark{\ref{Fermilab}},
Alan Uomoto\altaffilmark{\ref{JHU},\ref{CarnegieObs}},
Daniel E. Vanden Berk\altaffilmark{\ref{PSU}},
Jan Vandenberg\altaffilmark{\ref{JHU}},
S. Vidrih\altaffilmark{\ref{Cambridge}},
Michael S. Vogeley\altaffilmark{\ref{Drexel}},
Wolfgang Voges\altaffilmark{\ref{MPIEP}},
Nicole P. Vogt\altaffilmark{\ref{NMSU}},
David H. Weinberg\altaffilmark{\ref{OSU}},
Andrew A. West\altaffilmark{\ref{Berkeley}},
Simon D.M. White\altaffilmark{\ref{MPA}},
Brian Wilhite\altaffilmark{\ref{Illinois}},
Brian Yanny\altaffilmark{\ref{Fermilab}},
D. R. Yocum\altaffilmark{\ref{Fermilab}},
Donald G. York\altaffilmark{\ref{Chicago},\ref{EFI}},
Idit Zehavi\altaffilmark{\ref{Case}},
Stefano Zibetti\altaffilmark{\ref{MPIEP}},
Daniel B. Zucker\altaffilmark{\ref{MPIA},\ref{Cambridge}}
}

\altaffiltext{1}{
Fermi National Accelerator Laboratory, P.O. Box 500, Batavia, IL 60510.
\label{Fermilab}}

\altaffiltext{2}{
Department of Astronomy, University of Washington, Box 351580, Seattle, WA
98195.
\label{Washington}}

\altaffiltext{3}{
Department of Physics and Astronomy, University of Wyoming, Laramie, WY 82071.
\label{Wyoming}}

\altaffiltext{4}{
Apache Point Observatory, P.O. Box 59, Sunspot, NM 88349.
\label{APO}}

\altaffiltext{5}{
Department of Astrophysical Sciences, Princeton University, Princeton, NJ
08544.
\label{Princeton}}

\altaffiltext{6}{
Max-Planck-Institut f\"ur Astronomie, K\"onigstuhl 17, D-69117 Heidelberg,
Germany.
\label{MPIA}}

\altaffiltext{7}{
Astrophysics Research Institute,
Liverpool John Moores University,
Twelve Quays House, Egerton Wharf,
Birkenhead CH41 1LD, UK
\label{LJMU}}

\altaffiltext{8}{
Center for Astrophysical Sciences, Department of Physics and Astronomy, Johns
Hopkins University, 3400 North Charles Street, Baltimore, MD 21218. 
\label{JHU}}

\altaffiltext{9}{
Department of Physics \& Astronomy and Joint
Institute for Nuclear Astrophysics, Michigan State
University, East Lansing, MI 48824-1116
\label{MSUJINA}
}

\altaffiltext{10}{
Institute of Astronomy, University of Cambridge, Madingley Road,
Cambridge CB3 0HA, UK
\label{Cambridge}}

\altaffiltext{11}{
Center for Cosmology and Particle Physics,
Department of Physics,
New York University,
4 Washington Place,
New York, NY 10003.
\label{NYU}}

\altaffiltext{12}{
Department of Physics and Astronomy, University of Pennsylvania,
Philadelphia, PA 19104. 
\label{Penn}}

\altaffiltext{13}{
Centro de Astrof{\'\i}sica da Universidade do Porto, Rua 
das Estrelas - 4150-762 Porto, Portugal.
\label{Porto}}

\altaffiltext{14}{
Department of Astronomy
University of Illinois
1002 West Green Street, Urbana, IL 61801.
\label{Illinois}}

\altaffiltext{15}{Institut 
d'Estudis Espacials de Catalunya/CSIC, Gran Capit\'a 2-4,
E-08034 Barcelona, Spain.
\label{Barcelona}}

\altaffiltext{16}{
Department of Physics and Astronomy, University of Pittsburgh, 3941 O'Hara
Street, Pittsburgh, PA 15260.
\label{Pitt}}

\altaffiltext{17}{
Steward Observatory, 933 North Cherry Avenue, Tucson, AZ 85721.
\label{Arizona}}

\altaffiltext{18}{
Department of Astronomy and Astrophysics, University of Chicago, 5640 South
Ellis Avenue, Chicago, IL 60637.
\label{Chicago}}

\altaffiltext{19}{
Kavli Institute for Cosmological Physics, The University of Chicago,
5640 South Ellis Avenue, Chicago, IL 60637.
\label{CfCP}}

\altaffiltext{20}{
Department of Physics of Complex Systems, 
E\"{o}tv\"{o}s Lor\'and University, Pf.\ 32,
H-1518 Budapest, Hungary.
\label{Eotvos}}

\altaffiltext{21}{Institute of Astronomy,
School of Science, University of Tokyo,
Osawa 2-21-1, Mitaka, 181-0015, Japan.
\label{IoaUT}}

\altaffiltext{22}{
Space Telescope Science Institute, 3700 San Martin Drive, Baltimore, MD
21218.
\label{STScI}}

\altaffiltext{23}{Institute for Cosmic Ray Research, University of Tokyo, 5-1-5 Kashiwa,
 Kashiwa City, Chiba 277-8582, Japan.
\label{ICRRUT}}

\altaffiltext{24}{
Microsoft Research, 455 Market Street, Suite 1690, San Francisco, CA 94105.
\label{Microsoft}}

\altaffiltext{25}{
Astronomical Institute of the University of Basel, 
Department of Physics and Astronomy, Venusstrasse 7, CH-4102 Basel,
Switzerland
\label{Basel}}


\altaffiltext{26}{
Dept. of Physics \& Astronomy,
York University,
4700 Keele St.,
Toronto, ON, M3J 1P3,
Canada
\label{York}}

\altaffiltext{27}{
Institute for Astronomy and Computational Sciences
     Physics Department
     Catholic University of America
     Washington DC 20064
\label{Catholic}}

\altaffiltext{28}{
US Naval Observatory, 3540 Massachusetts Avenue NW, Washington, DC 20392.
\label{USNO}}

\altaffiltext{29}{
Code 7215, Remote Sensing Division
Naval Research Laboratory
4555 Overlook Avenue SW
Washington, DC 20392.
\label{NRL}}

\altaffiltext{30}{
Institute for Advanced Study
Einstein Drive
Princeton, NJ 08540
\label{IAS}}

\altaffiltext{31}{
Department of Astronomy, MSC 4500, New Mexico State University,
P.O. Box 30001, Las Cruces, NM 88003.
\label{NMSU}}

\altaffiltext{32}{National Astronomical Observatory, 
2-21-1 Osawa, Mitaka, Tokyo 181-8588, Japan.
\label{NAOJ}}

\altaffiltext{33}{
Astronomical Institute, Tohoku University,
Aoba, Sendai 980-8578, Japan
\label{Tohoku}}

\altaffiltext{34}{
School of Physics and Astronomy,
University of Southampton,
Southampton SO17 1BJ,
United Kingdom
\label{Southampton}}

\altaffiltext{35}{
Jet Propulsion Laboratory, 4800 Oak Drive, Pasadena, CA 91109
\label{JPL}}

\altaffiltext{36}{
California Institute of Technology, 1200 East California Blvd, Pasadena, CA
91125
\label{Caltech}}

\altaffiltext{37}{
ISR-4, MS D448, 
Los Alamos National Laboratory, P.O.Box 1663, Los Alamos, NM 87545.
\label{LANL}}

\altaffiltext{38}{
Max Planck Institut f\"ur Astrophysik, Postfach 1, 
D-85748 Garching, Germany.
\label{MPA}}

\altaffiltext{39}{
Subaru Telescope, 650 N. A'ohoku Place, Hilo, HI 96720, USA
\label{Subaru}}

\altaffiltext{40}{
Obserwatorium Astronomiczne na Suhorze, Akademia Pedogogiczna w
Krakowie, ulica Podchor\c{a}\.{z}ych 2,
PL-30-084 Krac\'ow, Poland.
\label{MSO}}

\altaffiltext{41}{
Enrico Fermi Institute, University of Chicago, 5640 South Ellis Avenue,
Chicago, IL 60637.
\label{EFI}}

\altaffiltext{42}{
Lawrence Berkeley National Laboratory, One Cyclotron Road,
Berkeley CA 94720.
\label{LBL}}

\altaffiltext{43}{
Gatan Inc., Pleasanton, CA 94588
\label{Gatan}}

\altaffiltext{44}{
Department of Physics, University of Chicago, 5640 South
Ellis Avenue, Chicago, IL 60637.
\label{ChicagoPhys}}

\altaffiltext{45}{
Astronomy Centre, University of Sussex, Falmer, Brighton BN1 9QJ, UK. 
\label{Sussex}}

\altaffiltext{46}{
Department of Astronomy \& Astrophysics, University of California, Santa
Cruz, CA 95064 
\label{SantaCruz}}

\altaffiltext{47}{
Joseph Henry Laboratories, Princeton University, Princeton, NJ
08544.
\label{Princetonphys}}

\altaffiltext{48}{
Instituto de Astrofisica de Canarias, La Laguna, Spain
\label{IAC}}

%

\altaffiltext{49}{
Department of Physics and Astrophysics,
 Nagoya University,
 Chikusa, Nagoya 464-8602,
 Japan.
\label{Nagoya}}

\altaffiltext{50}{
AOT-IC, MS H820, 
Los Alamos National Laboratory, P.O.Box 1663, Los Alamos, NM 87545.
\label{LANL2}}

\altaffiltext{51}{
Department of Physics, University of Michigan, 500 East University Avenue, Ann
Arbor, MI 48109.
\label{Michigan}}

\altaffiltext{52}{
SUPA; Institute for Astronomy,
Royal Observatory,
University of Edinburgh,
Blackford Hill,
Edinburgh EH9 3HJ,
UK.
\label{Edinburgh}}

\altaffiltext{53}{
US Naval Observatory, 
Flagstaff Station, 10391 W. Naval Observatory Road, Flagstaff, AZ
86001-8521.
\label{NOFS}}

\altaffiltext{54}{
Department of Physics, Applied Physics, and Astronomy, Rensselaer
Polytechnic Institute, 110 Eighth Street, Troy, NY 12180. 
\label{RPI}}

\altaffiltext{55}{
Institute of Cosmology and Gravitation (ICG),
Mercantile House, Hampshire Terrace,
Univ. of Portsmouth, Portsmouth, PO1 2EG, UK.
\label{Portsmouth}}

\altaffiltext{56}{
Gemini Observatory, 670 N. A'ohoku Place, Hilo, HI 96720, USA
\label{Gemini}}

%

\altaffiltext{57}{Department of 
Astronomy and Research Center for the Early Universe, 
University of Tokyo,
 7-3-1 Hongo, Bunkyo, Tokyo 113-0033, Japan.
\label{DoAUT}}

\altaffiltext{58}{
Korea Institute for Advanced Study,
207-43 Cheong-Yang-Ni, Dong-Dae-Mun-Gu
Seoul 130-722, Korea
\label{KIAS}}

\altaffiltext{59}{
FNRS
Institut  d'Astronomie et d'Astrophysique,
 Universit\'e Libre de Bruxelles, CP. 226, Boulevard du Triomphe, B-1050
 Bruxelles, Belgium.
\label{Bruxelles}}

\altaffiltext{60}{
Department of Physics, 
Drexel University, 3141 Chestnut Street, Philadelphia, PA 19104.
\label{Drexel}}

\altaffiltext{61}{
Department of Physics, Rochester Institute of Technology, 84 Lomb Memorial
Drive, Rochester, NY 14623-5603.
\label{RIT}}

\altaffiltext{62}{
UCO/Lick Observatory, University of California, Santa Cruz, CA 95064.
\label{Lick}}

\altaffiltext{63}{
Department of Astronomy and Astrophysics, 525 Davey Laboratory, 
Pennsylvania State
University, University Park, PA 16802.
\label{PSU}}

\altaffiltext{64}{
Department of Physics and Astronomy, Austin Peay State University,
P.O. Box 4608, Clarksville, TN 37040
\label{APeay}}

%
%

\altaffiltext{65}{
Adler Planetarium and Astronomy Museum,
1300 Lake Shore Drive,
Chicago, IL 60605.
\label{Adler}}

\altaffiltext{66}{
Department of Physics, The University of Tokyo, Tokyo 113-0033, Japan
\label{TokyoPhys}}

\altaffiltext{67}{
Institute for Astronomy, 2680 Woodlawn Road, Honolulu, HI 96822.
\label{Hawaii}}

\altaffiltext{68}{
Dept. of Physics, Massachusetts Institute of Technology, Cambridge,  
MA 02139.
\label{MIT}}

\altaffiltext{69}{
Observatories of the Carnegie Institution of Washington, 
813 Santa Barbara Street, 
Pasadena, CA  91101.
\label{CarnegieObs}}

\altaffiltext{70}{
Max-Planck-Institut f\"ur extraterrestrische Physik, 
Giessenbachstrasse 1, D-85741 Garching, Germany.
\label{MPIEP}}

\altaffiltext{71}{
Department of Astronomy, 
Ohio State University, 140 West 18th Avenue, Columbus, OH 43210.
\label{OSU}}

\altaffiltext{72}{
Astronomy Department, 601 Campbell Hall, University of
California, Berkeley, CA 94720-3411
\label{Berkeley}}

\altaffiltext{73}{
Department of Astronomy, Case Western Reserve University,
Cleveland, OH 44106
\label{Case}}
%
%

\shorttitle{SDSS DR5}
\shortauthors{Adelman-McCarthy \etal}

\begin{abstract}
This paper describes the Fifth Data Release (DR5) of the Sloan Digital Sky
Survey (SDSS). DR5 includes all survey quality data taken through June
2005 and represents the completion of the SDSS-I project
(whose successor, SDSS-II will continue through mid-2008).
It includes five-band photometric data for 217 million
objects selected over 8000 deg$^2$, and 1,048,960 spectra of
galaxies, quasars, and stars selected from 5713 deg$^2$ of
that imaging data.
These numbers represent a roughly 20\% increment over those of the
Fourth Data Release; all the data from previous data releases are
included in the present release.  
In addition to ``standard'' SDSS observations,
DR5 includes repeat scans of the 
southern equatorial stripe, imaging scans across M31 and the
core of the Perseus cluster of galaxies, and the first spectroscopic
data from SEGUE, a survey to explore the kinematics and 
chemical evolution of the Galaxy.
The catalog database incorporates several new features, including
photometric redshifts of galaxies, tables of matched objects in
overlap regions of the imaging survey, and tools that allow precise
computations of survey geometry for statistical investigations.
\end{abstract}

Submitted to The Astrophysical Journal Supplement Series, October 12, 2006

\section{Introduction}
\label{sec:introduction}

The primary goals of the Sloan Digital Sky Survey (SDSS) are:
a large-area, well-calibrated imaging survey of the north Galactic cap,
repeat imaging of an equatorial stripe in the south Galactic cap
to allow variability studies and deeper co-added imaging, and
spectroscopic surveys of well-defined samples of
roughly $10^6$ galaxies and $10^5$ quasars
(York \etal\ 2000).  The survey uses a dedicated, wide-field, 2.5m
telescope (Gunn \etal\ 2006) at Apache Point Observatory, New Mexico.
Imaging is carried out in drift-scan mode using a 142 mega-pixel
camera (Gunn \etal\ 1998) that gathers data in five broad bands,
$u\,g\,r\,i\,z$, 
spanning the range from 3000 to 10,000 \AA\ (Fukugita \etal\ 1996),
with an effective exposure time of 54.1 seconds per band.
The images are processed using specialized software (Lupton \etal\
2001; Stoughton \etal\ 2002; Lupton 2005), 
and are astrometrically (Pier
\etal\ 2003) and photometrically (Hogg \etal\ 2001; 
Tucker \etal\ 2006) calibrated using
observations of a set of primary standard stars (Smith \etal\ 2002)
observed on a neighboring 20-inch telescope.  

Objects are selected from the imaging data for spectroscopy using a
variety of algorithms, including a complete sample of galaxies with
Petrosian (1976) 
$r$ magnitudes brighter than 17.77 (Strauss \etal\ 2002), a deeper
sample of color- and magnitude-selected luminous
red galaxies (LRGs) from redshift 0.15 to beyond 0.5 (Eisenstein \etal\
2001), a color-selected sample of quasars with $0 < z < 5.5$
(Richards \etal\ 2002), optical counterparts to ROSAT X-ray
sources (Anderson \etal\ 2003), and a variety of stellar and
calibrating objects (Stoughton \etal\ 2002; Adelman-McCarthy \etal\ 2006).  
These targets are observed by a pair of double spectrographs fed
by 640 optical fibers, each $3''$ in diameter, 
plugged into aluminum plates 2.98$^\circ$ in diameter.
The resulting spectra cover the wavelength range $3800-9200$ \AA\ with
a resolution of $\lambda/\Delta \lambda \approx 2000$.  
The finite size of the fiber cladding means
that only one of two objects closer than $55''$ can be targeted on a
given plate; this restriction results in a roughly 10\%
incompleteness in galaxy spectroscopy, but this incompleteness is
well characterized and is generally straightforward to correct in statistical
calculations (e.g., Zehavi et al.\ 2002).

This paper presents the Fifth Data Release (DR5) of the SDSS,
which follows the Early Data Release of commissioning data
(EDR; Stoughton \etal\ 2002) and the regular data releases
DR1-DR4 (Abazajian \etal\ 2003, 2004, 2005;
Adelman-McCarthy \etal\ 2006).
These data releases are cumulative, so all observations in the earlier
releases are also included in DR5.  
There have been no substantive changes to the imaging or
spectroscopic software since DR2, so DR5 includes 
data identical to DR2-DR4 in the overlapping regions.
Finkbeiner \etal\ (2004) presented a separate
(``Orion'') release of imaging data outside the formal SDSS footprint,
mostly at low Galactic latitudes.

The Fifth Data Release includes all survey quality data that were
taken as part of ``SDSS-I,'' the phase of the SDSS that ran through
June 2005, including a variety of imaging scans and spectroscopic
observations taken outside of the standard survey footprint or
with non-standard spectroscopic target selection.
The second ``SDSS-II'' phase, which includes a number of new
participating institutions and will continue through mid-2008,
consists of three distinct 
surveys: the Sloan Legacy Survey, the Sloan Supernova Survey,
and the Sloan Extension for Galactic Understanding and Exploration
(SEGUE).  The Legacy survey is essentially a continuation of SDSS-I,
with the goal of completing imaging and spectroscopy over about
8000 deg$^2$ of the north Galactic cap.
The Supernova Survey (J. Frieman \etal\ 2007, in preparation)
repeatedly scans a 300 square degree area in
the south Galactic cap during the fall months to detect and measure
time variable objects, especially Type Ia supernovae 
(out to $z\approx 0.4$)
that can
be used to measure the cosmic expansion history.
SEGUE includes 3500 deg$^2$ of new imaging, mostly at 
Galactic latitudes below those of the original SDSS footprint,
and spectroscopy of about 240,000 selected stellar targets to
study the structure, chemical evolution, and stellar content of
the Milky Way.  Future SDSS data releases will include data from
all three surveys, and some early data from SEGUE are included
in DR5.  An initial release of imaging data and uncalibrated
object catalogs from the Autumn 2005 season of the Supernova Survey
is available at {\tt http://www.sdss.org/drsn1/DRSN1\_data\_release.html},
but it is not part of DR5.

Section~\ref{sec:DR5} of this paper describes the contents
of DR5, and \S\ref{sec:quality} summarizes information about
data quality, including new tests of spectrophotometric accuracy.
Section~\ref{sec:features} describes several new features of DR5:
photometric redshifts for galaxies, 
``sector/region'' tables
for precisely defining the survey geometry, 
and tools for matching repeat observations of the same objects.
We conclude in \S~\ref{sec:conclusions}.  

\section{What is included in DR5}
\label{sec:DR5}

As described by Stoughton \etal\ (2002), public SDSS data are
available both as flat files (from the Data Archive Server, or DAS) 
and via a flexible web interface to the SDSS database (the 
Catalog Archive Server, or CAS).  Information about and entry 
points to both interfaces can be found at {\tt http://www.sdss.org/dr5}.  
The CAS is a convenient and powerful tool for selecting objects
found in the SDSS based on their location, photometric parameters,
and (if they were observed spectroscopically) spectroscopic parameters.
FITS images and spectra for individual objects and fields
are available from the CAS; the DAS should be used for bulk
downloads of large quantities of data.
Links to extensive documentation and examples
are available on the above web site.

\begin{figure}
\plotone{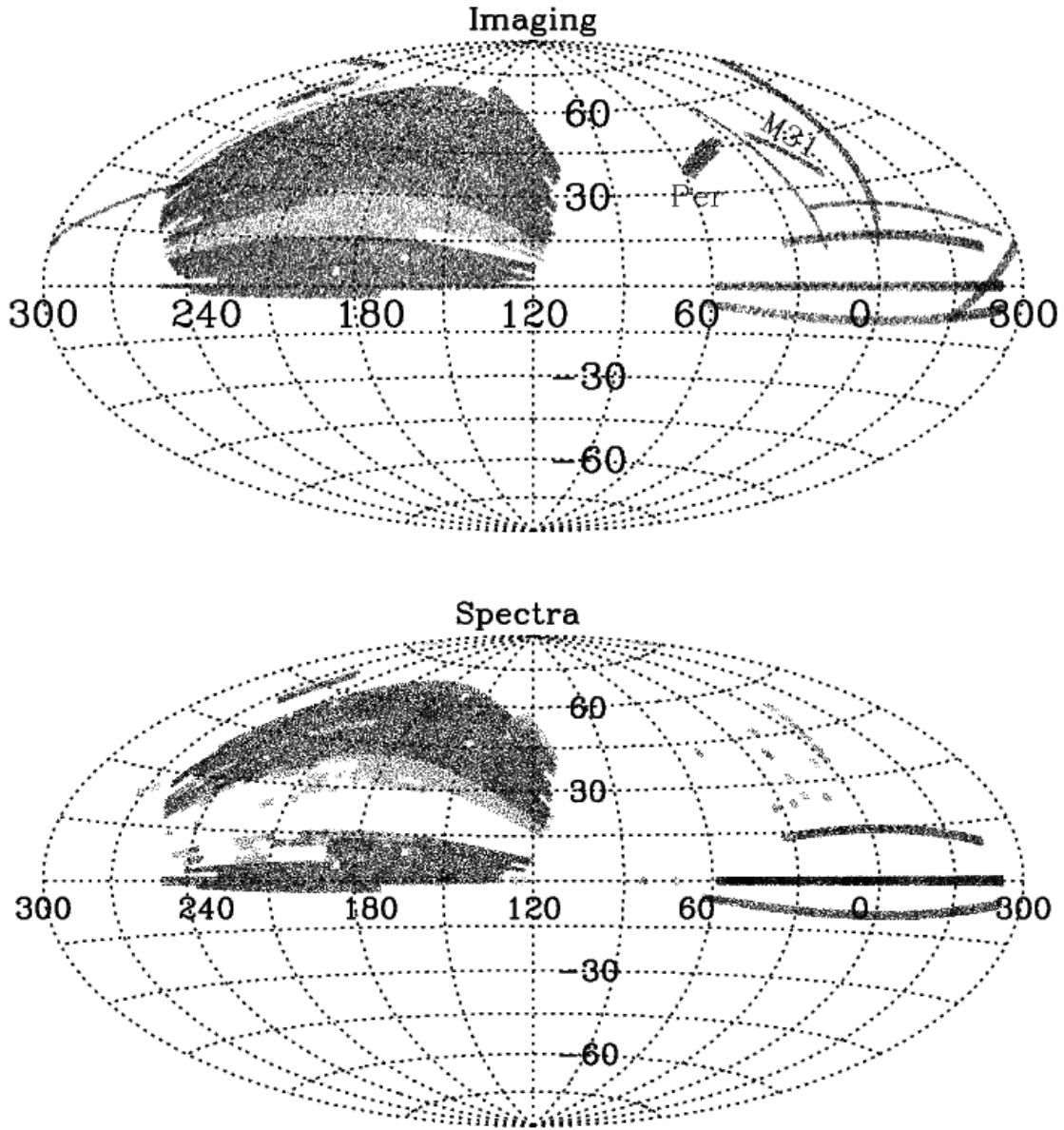}
\caption{
The distribution on the sky of SDSS imaging (upper panel) and
spectroscopy (lower panel) included in DR5, shown in J2000 equatorial
coordinates.  The regions of sky that are new to DR5 are shaded more lightly.
The upper panel includes both those regions included in the CAS
(totaling 8000 deg$^2$) and the supplementary imaging runs
available only through the DAS, which consist of SEGUE scans
at low Galactic latitude and scans through M31 and the Perseus cluster.
\label{fig:skydist}}\end{figure}

The principal SDSS imaging data are taken
along a series of great-circle stripes that aim to fill
a contiguous area in the north Galactic cap, and along three
non-contiguous stripes in the south Galactic cap.  
Each filled stripe consists of two interleaved strips
because of the gaps between columns of CCDs
in the imaging camera (see Gunn \etal\ 1998; York \etal\ 2000).
Figure~\ref{fig:skydist} shows the region of sky included in DR5,  
in imaging (top) and spectroscopy (bottom).  In contrast to DR4,
the imaging available in DR5 covers an essentially contiguous region of
the north Galactic cap, with 
a few small patches totaling $\sim 200$ square degrees
remaining (nearly all of this area will be included in DR6).
The area covered by the DR5 primary
imaging survey (including the southern stripes but not counting
these patches) is 8000 deg$^2$.
The great circle
stripes in the north overlap at the poles of the survey; 
21\% of this region of sky is covered more than once. 
In any region where imaging runs overlap, one run is
declared primary and used for spectroscopic target
selection, and other runs are declared secondary.  DR5 includes 
both the primary and secondary (repeat) observations
of each area and source (see \S\ref{sec:match}).

As spectroscopic observations necessarily
lag the imaging, the DR5 spectroscopic area still has the gap
at intermediate declinations that was present in the DR4 imaging
coverage.  The area covered by the spectroscopic
survey is 5713 deg$^2$.
The spectroscopic data include 1,048,960 spectra,
arrayed on 1639 plates of 640 fibers each.  
Thirty-two fibers per plate are
devoted to measurements of sky. 
Automated spectral classification yields
approximately 675,000 galaxies,
90,000 quasars, and 216,000 stars.
Nearly 99\% of all spectra are of high enough quality to yield an
unambiguous classification and redshift; most of the unidentified targets
are either faint ($r>20$) or have featureless spectra
(hot stars or blazar-like AGN; see Collinge \etal\ 2005).  
However, in rare cases the assigned redshift is far from the
true redshift, so for an object with unusual
properties it is important to
examine the spectra and 
to check for flags that can indicate data quality or classification
problems.
As described in the DR4 paper (Adelman-McCarthy \etal\ 2006),
a number of plates have duplicate
observations, usually just one but in some cases several.
DR5 includes 62 duplicates of 53 unique 
main survey plates, and ten duplicates of special plates
which take spectra outside the standard survey target
selection.
Some main-survey objects are also reobserved on adjacent
plates to check the end-to-end reproducibility of spectroscopy.
In total, about 2\% of main-survey objects have one or more repeat spectra.

\begin{figure}
\plotone{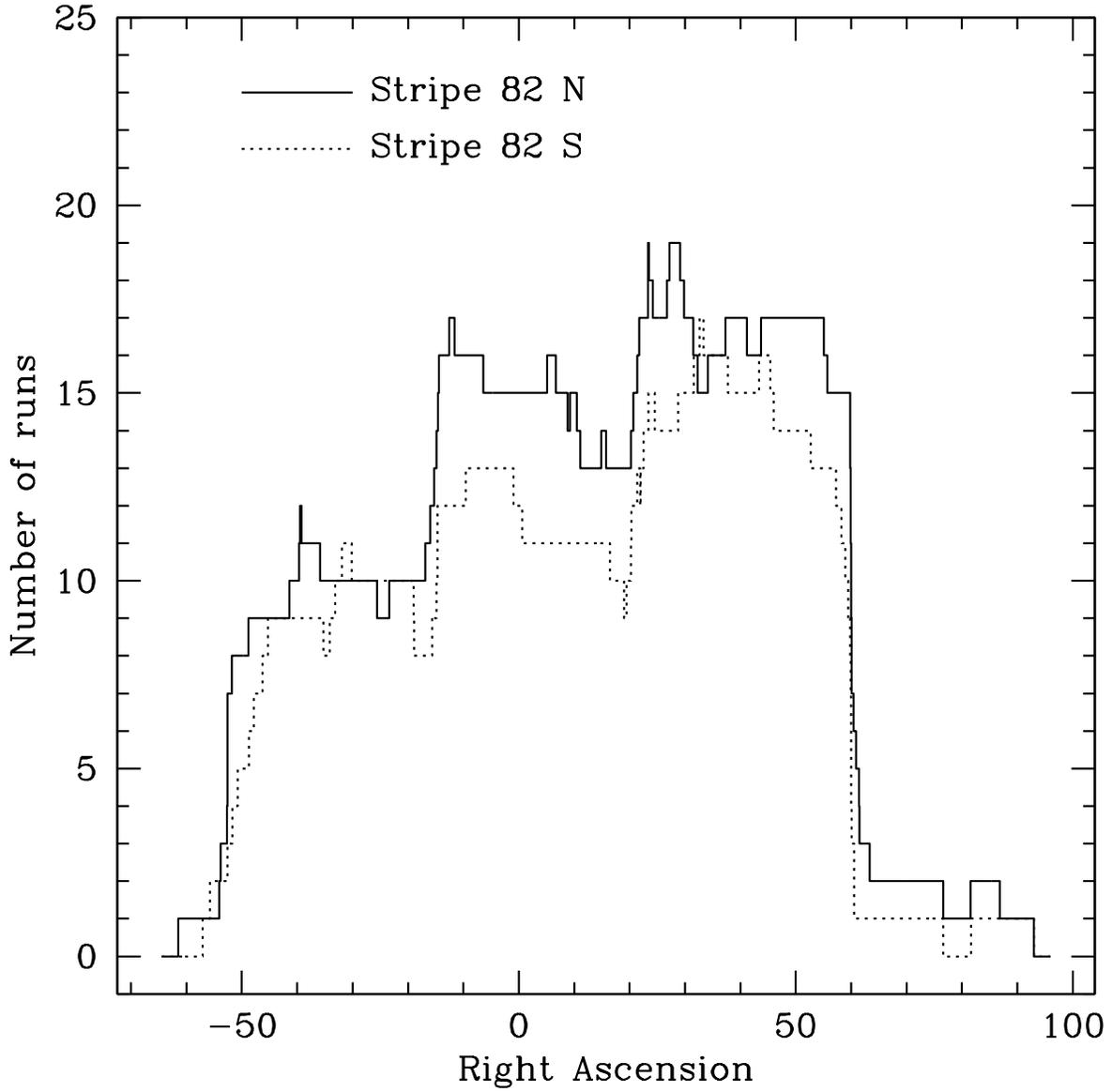}
\caption{
Coverage of the southern equatorial stripe in DR5.
Solid and dotted lines show the number of photometric runs
covering regions of different right ascension, for the
northern and southern strips, respectively.
\label{fig:stripe82}}
\end{figure}

In the Fall months, when the southern Galactic cap is visible in the
northern hemisphere, the SDSS imaging has been confined to a stripe
along the Celestial Equator, plus two ``outrigger'' stripes,
centered roughly at $\delta = +15^\circ$ and $\delta = -10^\circ$,
respectively (these are visible on the right-hand-side of the 
panels of Figure~\ref{fig:skydist}).  We have performed multiple
imaging passes of the southern equatorial stripe
(a.k.a. Stripe 82, spanning $\rm 22^h\,20^m < \alpha < 3^h\,20^m$, 
$-1.25^\circ < \delta < + 1.25$, in J2000 coordinates), 
which can be used for variability studies and for 
co-addition to create deeper summed images.
Previous data releases have included only a single epoch of these
observations.  In DR5, we make available 36 runs on the northern
strip of this stripe and 29 runs on the southern strip; these are
all the observations of Stripe 82 carried out before July 2005 that are
of survey quality. 
Each individual run covers only part of the full right
ascension range of the stripe; Figure~\ref{fig:stripe82} shows the number
of passes available along the northern and southern strips, as a
function of right ascension.  
The central regions of the stripe have typically been covered 10-20
times.  The extra runs are available in DR5
only through the DR supplemental DAS, described at
{\tt http://www.sdss.org/dr5/start/aboutdr5sup.html}. In future data releases, 
they will be made 
available through the CAS as well.  
Note that DR5 does not include
those runs on Stripe 82 at larger right ascension, in the region of
Orion, as described by Finkbeiner \etal\ (2004).  Those runs
continue to be made available through the websites indicated in that
paper. 

\begin{figure}
\epsscale{0.5}
\plotone{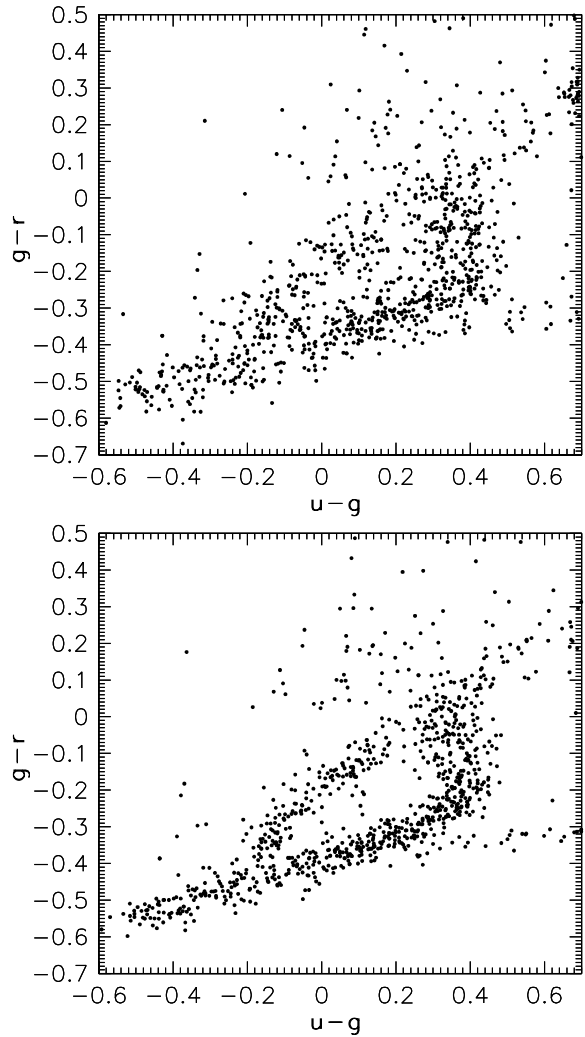}
\caption{
The $g-r$ vs. $u-g$ color-color diagram for the blue, non-variable
point sources with $u<20$ in the equatorial stripe 
(from Ivezi\'c \etal [2007]).
The top panel shows results
using single-epoch DR5 photometry, while the bottom panel shows
the striking improvement obtained by averaging the photometric
measurements from all of the imaging passes, allowing clear
separation between the sequences of helium white dwarfs
(the top side of the ``triangle'') and hydrogen white dwarfs
(which lie along the other two sides).  
This region of color
space also includes white dwarf--M dwarf pairs, hot subdwarfs,
and quasars (see, e.g., the discussion of Eisenstein et al.\ [2006]).
Main sequence and red giant stars (far more numerous, of course),
are mostly off the diagram to the upper right.
\label{fig:colorcolor}}
\end{figure}

A combined, deep image of the full equatorial stripe is being prepared
and will be made available in a future data release.
However, for objects that can be detected in a single pass, 
the benefits of co-addition can mostly be realized simply
by averaging the photometric measurements from the multiple
passes, using the multiple entries in the photometric catalog rather 
than analyzing a summed image.  Figure~\ref{fig:colorcolor},
based on the Stripe 82 stellar catalog of 
Ivezi\'c \etal\ (2007),
demonstrates this improvement, 
showing the $g-r$ vs. $u-g$ color-color diagram for blue,
non-variable point sources (mostly white dwarfs) 
in Stripe 82.  
Data co-added at the catalog level have been used to search
for faint quasars (Jiang \etal\ 2006), to measure the dispersion
in galaxy colors on the red sequence (Cool \etal\ 2006),
and to improve the signal-to-noise ratio of galaxy 
$u$-band Petrosian magnitudes (Baldry \etal\ 2005).
The Stripe 82 data have also 
been used to search for variable and high proper motion objects
(e.g., Ivezi\'c \etal\ 2003) and to test the covariance
of photometric errors among bands and among multiple objects
in the same fields (Scranton \etal\ 2005).
Because the catalogs from the multiple Stripe 82 scans are not
yet available in the CAS, averaging or variability searches must
be done by downloading object tables from the DAS and 
identifying repeat observations of the same object by 
positional matching.

In addition to the repeat scans on Stripe 82, several
imaging runs outside of the standard footprint are included:
\begin{itemize} 
\item Two runs that together make a $2.5^\circ$ 
  stripe crossing M31, the Andromeda Galaxy.  These imaging data have
  been used to search for substructure in M31's halo (e.g., Zucker
  \etal\ 2004ab).  
 \item Five runs that together cover 78 deg$^2$ centered roughly on
   the low-redshift Perseus cluster of galaxies. 
 \item Ten runs of imaging data taken as part of the SEGUE survey,
	including stripes at $l=50^\circ\>$ ($-46^\circ < b < -8^\circ$),
	$l=110^\circ\>$ ($-36^\circ < b < 29.5^\circ$), and
	$l=130^\circ\>$ ($-49 < b < -18.6$), and a stripe that
	runs for 20 degrees along $\delta \approx 25^\circ$.
\end{itemize}
As with the repeat scans of Stripe 82, objects detected in these
runs are recorded in the DRsupplemental DAS: 
{\tt http://www.sdss.org/dr5/start/aboutdrsup.html}, but they are not, as yet, 
available in the CAS.
All these runs are in quite crowded fields, as they tend to go to low
Galactic latitude, or pass through the center of M31.  The completeness
and accuracy of the photometry produced by the automated 
SDSS pipeline becomes suspect in crowded fields, so these data should be used
with care.  
Plots and tables of the field-by-field data
quality for these runs may be accessed at 
{\tt http://das.sdss.org/DRsup/data/imaging/QA/summaryQA\_analyzePC.html}.

Because of the relatively small footprint of the imaging in the southern
Galactic cap, the spectroscopy of targets selected by
our normal algorithms was completed
quite early in the survey; most of these data
were included already in DR1.  We generally restrict imaging observations
to pristine conditions, when the moon is below the horizon, the sky is
cloudless, 
and the seeing is good.  To make optimal use of the remaining time,
we undertook a series of 
spectroscopic observing programs, based mostly on the imaging data
of the equatorial stripe in the southern Galactic cap, designed to
go beyond the science goals of the main survey.  DR5 includes 299
plates from these programs, 
carried out in the Fall months of 2001--2004,  
with a total of 204,160 spectra.
The great majority of these plates were already included in DR4;
the target selection for them is described in the DR4 paper
(Adelman-McCarthy \etal\ 2006), and we will not repeat it here.
The science objectives include
studies of galactic kinematics, calibration of photometric redshifts,
evaluation of the completeness of the quasar survey
(Vanden Berk \etal\ 2005),
and surveys of galaxies that fall outside of the standard
survey selection criteria (Baldry \etal\ 2005).  

DR5 includes a total of 84 special plates that were not included
in DR4.  All of these were obtained as early data of the SEGUE program.
Each SEGUE pointing includes two 640-fiber plates of different
exposure times, with 592 brighter
($13 < g < 18$) and 560 fainter ($18 < g < 20$) stars targeted.
The remaining targets are calibration standards and sky fibers.
Target selection algorithms, 
which are outlined in 
Adelman-McCarthy \etal\ (2006) and will be described
more fully in a future paper,
identify candidate stars in the following categories:
white dwarfs (25 per pointing), cool white dwarfs (10),
A/BHB stars (150), F turnoff and sub-dwarf stars (150),
G stars (375), K giants (100), low metallicity candidates (150),
K dwarfs (125), M dwarfs (50), and AGB candidates (10).
These plates are listed and described at
{\tt http://www.sdss.org/dr5/products/spectra/special.html}.


Tables~\ref{table:imaging} and~\ref{table:spectroscopy}
summarize the characteristics of the DR5 imaging and spectroscopic
surveys, respectively.
Note that the ``star'' and ``galaxy'' divisions in 
Table~\ref{table:imaging} refer to the 
photometric pipeline classifications; ``stars'' include 
quasars and any other unresolved sources, and
``galaxies'' are all resolved objects, including
airplane and satellite trails, etc.
Classifications in Table~\ref{table:spectroscopy} are those
returned by the spectroscopic pipeline; note, in particular,
that the ``quasar'' classification (based on the presence
of a securely detected, high excitation emission line with
FWHM broader than 1000~km~sec$^{-1}$) does not include
any explicit luminosity cut.

DR5 contains several QSO-related tables and views. The
{\tt QuasarCatalog} table lists the individually inspected, luminosity
and line-width restricted, bonafide quasars from the DR3 sample as
published by Schneider et al. (2005). A similar catalog is now
being created for DR5 (Schneider et al. 2007).  The
{\tt QSOBunch} table contains a record for each ``object'' flagged as a
potential QSO in any of three catalog tables: {\tt Target.PhotoObjAll},
{\tt Best.PhotoObjAll} or {\tt SpecObj}. In such cases a bunch record
describing the primary photo, target, and spectro objects within
1.5 arcseconds of that object is created.  Identifiers of nearby
objects from each catalog are combined into {\tt QSOConcordanceAll}
records that point to the {\tt QSOBunch} record.  Those identifiers
in turn point to the {\tt QSObest}, {\tt QSOtarget}, and {\tt QSOspec} 
tables that carry more detailed information about each object.
Thus, the {\tt QuasarCatalog} table provides straightforward
access to a set of carefully vetted quasars with well defined
selection criteria, while the {\tt QSOConcordanceAll} table
can be used to identify all objects that were flagged as potential
quasars based on photometry and/or spectroscopy.

\begin{deluxetable}{lr}
\tablecaption{Characteristics of the DR5 Imaging Survey
              \label{table:imaging}}

\startdata

\cutinhead{}

 Footprint area & 8000\ deg$^2$ (20\% increment over DR4)\\
 Imaging catalog & 217 million unique objects \\
 AB Magnitude limits:\tablenotemark{a} \\
\qquad $u$      & 22.0 mag\\
\qquad $g$      & 22.2 mag\\
\qquad $r$      & 22.2 mag\\
\qquad $i$      & 21.3 mag\\
\qquad $z$      & 20.5 mag\\
Median PSF width     & $1.4^{\prime\prime}$ in $r$ \\
 RMS photometric calibration errors: \\
\qquad   $r$ & 2\% \\
\qquad   $u-g$ & 3\% \\
\qquad   $g-r$ & 2\% \\
\qquad   $r-i$ & 2\% \\
\qquad   $i-z$ & 3\% \\
 Astrometry errors   & $< 0.1^{\prime\prime}$ rms absolute per coordinate \\
Object Counts:\tablenotemark{b}\\
\qquad Stars, primary	& 85,383,971\\
\qquad Stars, secondary & 28,201,858\\
\qquad Galaxies, primary   & 131,721,365\\
\qquad Galaxies, secondary & 33,044,047\\

\enddata

\tablenotetext{a}{95\% completeness for point sources in
typical seeing; 50\% completeness numbers are generally 0.4 mag
fainter.  The difference between ``asinh'' magnitudes and 
conventional magnitudes is $0.004 - 0.015$ at the 95\% limits
and $0.008 - 0.03$ at the 50\% limits, smaller than the uncertainty
in conversion of magnitudes between surveys used to estimate the
completeness.}
\tablenotetext{b}{Primary imaging objects are those in the primary
imaging area; secondary objects are in repeat imaging, so they
are typically repeats of primary objects.}
\end{deluxetable}

\begin{deluxetable}{lr}
\tablecaption{Characteristics of the DR5 Spectroscopic Survey
              \label{table:spectroscopy}}

\startdata
\cutinhead{\bf Main Survey}

Footprint area  & 5713\ deg$^2$ (19\% increment over DR4)\\
Wavelength coverage & 3800--9200\AA\\
 Resolution $\lambda/\Delta \lambda$    &1800--2100 \\
 Signal-to-noise ratio\tablenotemark{a} &$>4$ per pixel at $g=20.2$\\
 Wavelength calibration errors & $<5$ km sec$^{-1}$\\
 Redshift accuracy & 30 km sec$^{-1}$ rms for Main galaxies\\
                   & $\sim 99\%$ of classifications and redshifts are reliable\\
 Number of plates & 1639 \\
 Number of spectra\tablenotemark{b}& 1,048,960\\
 \qquad  Galaxies   &  674,741 \\
 \qquad\quad  Science primary galaxies&  561,530 \\
 \qquad  Quasars    & 90,596 \\
 \qquad\quad  Science primary quasars&  75,005\\
 \qquad  Stars      & 215,781 \\
 \qquad  Sky        & 55,555 \\
 \qquad  Unclassifiable & 12,287\\ 

\cutinhead{\bf Additional Spectroscopy}

  Repeat of main survey plates & 62 plates \\
  SEGUE and SEGUE test plates & 80 plates (2 repeated) \\
  Other southern programs & 219 plates (8 repeated) \\

\enddata
\tablenotetext{a}{Pixel size is 69 km$\,{\rm s}^{-1}$, 
varying from 0.9\AA\ (blue end)
to 2.1\AA\ (red end).}
\tablenotetext{b}{
Science primary objects define the set of unique science spectra
of objects from main-survey plates (i.e., they exclude repeat observations,
sky fibers, spectrophotometric standards, and objects from special plates).
}
\end{deluxetable}

\section{Data Quality}
\label{sec:quality}

\begin{figure}
\plotone{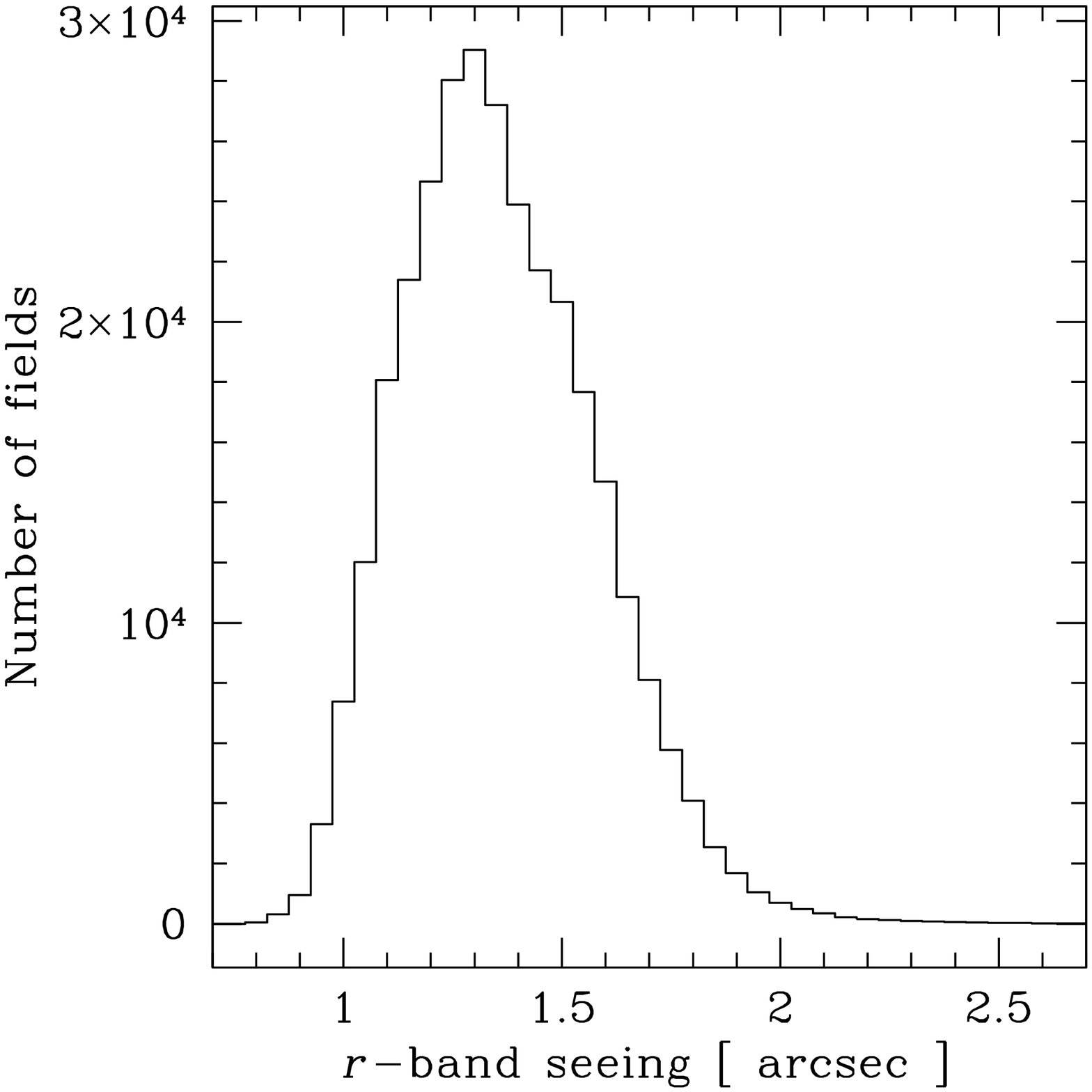}
\caption{
The distribution of image quality (FWHM of point sources)
in the imaging survey, measured in $r$ band.
\label{fig:seeing}}\end{figure}

SDSS imaging data are obtained under photometric conditions,
as determined by observations from the 0.5-m photometric monitoring
telescope and a 10$\mu$m ``cloud camera''
(Hogg \etal\ 2001; Tucker \etal\ 2006).
The median seeing of the imaging data is $1.4''$
in the $r$ band, and essentially all imaging data accepted
as survey quality have seeing better than $2''$ 
(see Figure~\ref{fig:seeing}).
The 95\% completeness limit for detection of point sources in
the $r$ band is 22.2 mag, estimated from comparison to deeper surveys
(COMBO-17 and CNOC-2).  
Constancy of stellar population colors shows that photometric
calibration over the survey area is
accurate to roughly 0.02 mag
in the $g, r$ and $i$ bands, and 0.03 mag in $u$ and
$z$ (Ivezi\'c \etal\ 2004).
Analysis of multiple observations of the southern Equatorial stripe
shows that photometry of bright stars is repeatable at better
than 0.01 mag in all bands and that the photometric pipeline
correctly estimates random photometric errors
(Ivezi\'c \etal 2007).
All magnitudes are roughly on an AB system
(Oke \& Gunn 1983) and use the ``asinh'' scale described by Lupton,
Gunn, \& Szalay (1999).  
The astrometric calibration precision
is better than $0.1''$ rms per coordinate
(Pier \etal\ 2003).  

The wavelength calibration uncertainty for SDSS spectra is roughly 0.05 \AA. 
Note that spectra in DR5 (and DR2-DR4) are {\it not} corrected for 
Galactic extinction; this is a change relative to DR1.
The spectra are flux-calibrated using observations of F subdwarfs,
which are targeted for this purpose on each spectroscopic plate;
the calibration procedure is described in \S 4.1 of
Abazajian et al.\ (2004).
Wilhite et al.\ (2005) discuss the repeatability of stellar spectra
taken more than 50 days apart. 
Their Figure 4 shows that the distribution of the fractional difference
from one observation to another in the flux summed over all pixels in
non-variable stars has a 68\% full-width of $\sim 5-8\%$, depending on
signal-to-noise ratio.  Their Figure 5 shows that the typical offset
in the calibration between two epochs of a single plate is $1-3$\%
over the full observed wavelength range, with no strong features at
any wavelength.

A useful way to test the quality of spectrophotometry on small
scales ($< 500$\AA) is to observe a population of identical
objects at a range of redshifts.  Spectrophotometric
residuals may then be computed by dividing the restframe spectra of
objects in different redshift bins.  While no ideal population
of identical objects
exists, elliptical galaxies have spectra that are similar, on average,
over the redshift range $z=0.04-0.20$, since they are no longer forming stars.

We select ellipticals for this test
using their position in the color-magnitude diagram,
with an additional cut on the H$\alpha$ equivalent width of 2\AA\ to
exclude any objects with
on-going star formation.  We average 300 to 1000 spectra in the
restframe in 160 bins of 0.001 in redshift from $z=0.04-0.20$.  To determine
the spectrophotometry residuals we must fit out any
evolution with redshift, which can arise from a combination of true
passive evolution, slight changes in sample selection, and aperture
effects.  This is done by fitting a fourth-order polynomial to flux as
function of redshift for each {\it rest-frame} wavelength.  We divide the
rest-frame spectra by these fits and interpolate back to the observed
frame.  The median of the residual spectra in the observed frame 
provides a measure of the spectrophotometry error, i.e., the mean
factor by which the flux-calibrated spectrum provided by the
spectroscopic pipeline is high or low compared to a perfectly
calibrated spectrum.  Since the evolutionary fits are themselves
affected by the spectrophotometry errors, we apply the estimated
correction to the averaged spectra and iterate the process,
which converges rapidly.

\begin{figure}
\plotone{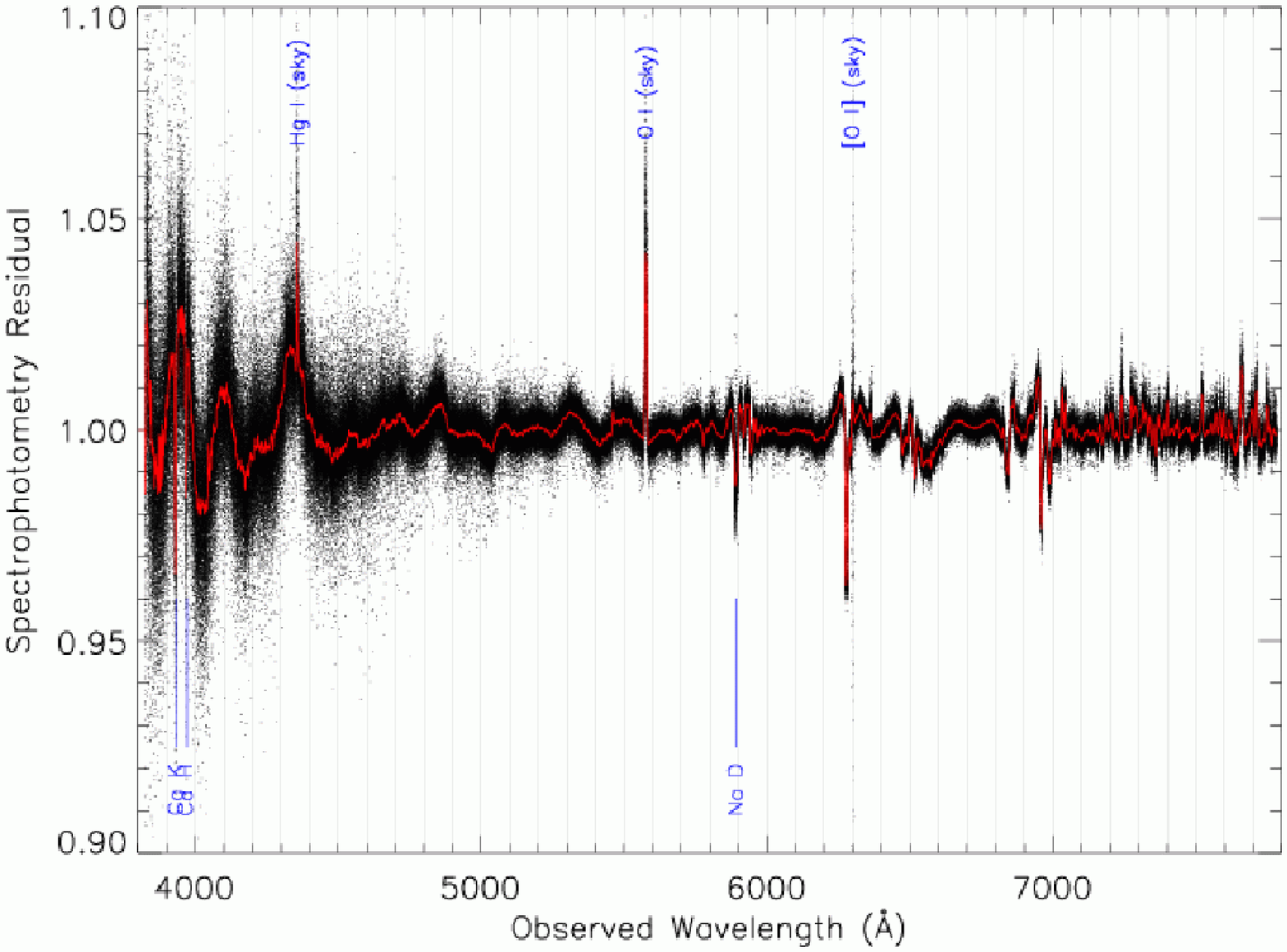}
\caption{
Test of spectrophotometric accuracy, performed by dividing the
rest-frame spectra of elliptical galaxies observed over the
redshift range $0.04 \leq z \leq 0.2$ (see text).  
Points show the residual inferred from 160 redshift-bin spectra (each an
average of 300-1000 individual galaxies) spaced by $\Delta z=0.01$,
and the central line shows the median residual.
\label{fig:specphoto}}\end{figure}

Figure~\ref{fig:specphoto} shows the spectrophotometry residuals
inferred from each of the 160 composite spectra, and the
median of these residuals.  There are sharp features associated with 
calcium and sodium absorption, probably originating in the 
Galactic interstellar medium, and with night sky emission lines.
The most worrisome features are the wiggles below 4500 \AA,
with amplitude of $\sim 3\%$,
centered on Ca H and K, H$\delta$, and H$\gamma$.  
The coincidence of these wiggles with known spectral features
suggests that these residuals
are caused by a systematic mismatch between the
spectrophotometric standard stars and the model F-stars used in the
calibration pipeline.  

One obvious question is the scale at which we can measure
spectrophotometry errors with this technique.  This scale is set by
our ability to discriminate evolution effects from the
spectrophotometry residuals, which in turn is related to the wavelength shift
between our high- and low-redshift bins.  We have tested the technique
empirically by
adding sine and cosine modulations with different periods to the observed 
frame and the seeing how well we recover them.  Residuals seem
to be well measured on scales less than 500\AA, i.e., 
Figure~\ref{fig:specphoto} should reveal any systematic errors
in SDSS photometry with periods shorter than this.
On larger scales, we must rely on the F star spectral models,
on tests against white dwarf model spectra
(see figure 4 of Abazajian \etal\ 2004),
and on checks of synthesized magnitudes against the photometry.
Collectively, these tests imply that the
flux-calibrated SDSS spectra can be used for spectrophotometry
at the few percent level.

\section{New Features of DR5}
\label{sec:features}

\subsection{Photometric Redshifts for Galaxies}
\label{sec:photoz}

DR5 includes two estimates of photometric redshifts for galaxies,
calculated with two independent techniques.\footnote{See
{\tt http://skyserver.elte.hu/PhotoZ/} and
{\tt http://yummy.uchicago.edu/SDSS/} for details.}
The first uses the template fitting
algorithm described by Csabai et al. (2003),
which compares the expected colors of 
a galaxy (derived from template spectral energy distributions) 
with those observed for an individual galaxy. 
A common approach
for template fitting is to take a small number of spectral templates T 
(e.g., E, Sbc, Scd, and Irr galaxies) and choose the best fit by 
optimizing the likelihood of the fit as a function of redshift, type, 
and luminosity, $p(z, T, L)$. We use a variant of this method that 
incorporates a
continuous distribution of spectral templates, enabling the error 
function in redshift and type to be well defined.
Since a representative set of photometrically calibrated spectra in the 
full wavelength range of the filters is not easy to obtain, we have 
started from the empirical templates of Coleman, Wu, \& Weedman (1980),
extended them with spectral synthesis models, and adjusted 
them to fit the colors of galaxies in the training set
(Budavari et al. 2000).
The results are listed in the CAS table {\tt Photoz}, which includes
the estimate of the redshift, spectral type, rest-frame colors,
rest-frame absolute magnitudes, errors on all of these quantities,
and a quality flag.  All photometric objects have an entry in 
the {\tt PhotoZ} table, regardless of whether they are photometrically
classified as galaxies or stars, so it is essential to consult
the quality flag and error characterizations when using the
photometric redshifts.

The second photometric redshift estimate
uses a neural network method that is very
similar in implementation to that of Collister \& Lahav (2004).
The training set consists of 140,000 single pass SDSS photometry
measurements with spectroscopic redshifts from various sources:
the SDSS (110,000 redshifts), CNOC2 (Yee \etal\ 2000; 9000 redshifts)
CFRS (Lilly \etal\ 1995; 1000 redshifts),
DEEP and DEEP2 (Weiner \etal\ 2005; 1700 redshifts),
TKRS/GOODS (Wirth \etal\ 2004; 300 redshifts),
and the 2SLAQ LRG survey (Cannon \etal\ 2006; 27,000 redshifts).
The SDSS portion of the training set consists of a representative
sampling of the SDSS Main, LRG, and southern survey spectroscopic data;
the other surveys are used to augment the training set at 
magnitudes fainter than probed by the SDSS spectroscopic samples.
Note that the training set multiply counts independent, repeat SDSS photometric 
measurements of the same objects, in particular on SDSS Stripe 82. 
Photometric redshift errors are computed using the Nearest Neighbor 
Error method (NNE), which assigns to each object an error based
on the photometric redshift error distribution of objects with 
similar magnitude and color in the training set
(for which the true redshifts are known), and this approach is
found to accurately predict the errors (H. Oyaizu \etal, in preparation).
The trained network is tested on a larger validation set consisting of 
1,700,000 objects with SDSS photometry (counting independent repeat
measurements) and for which spectroscopic redshifts are available.
The input catalogs for these photometric redshift measurements
were derived from the SDSS photo pipeline outputs, but with a few 
additional cuts employed to improve the star-galaxy separation,
using the PSF probability and the lensing smear polarizability 
(Sheldon \etal\ 2004).  The photometric sample was cut at a 
galaxy probability greater than 0.8, which is very stringent, and a
smear polarizability less than 0.8, and further cuts on magnitude were 
also made; hence not all DR5 objects are included.
The {\tt Photoz2} table lists a photometric redshift, an error, and 
a quality flag.  For objects with all five SDSS magnitudes
measured, the flag is set to 0 if $r \leq 20$ or 2 if $r > 20$;
photometric redshifts for flag $=2$ objects are subject to larger
uncertainties.  Objects not satisfying the above conditions have
flag set to 1 or 3 and their photometric redshifts should not be used.
There are 12.6 million objects in the DR5 data set with a
{\tt Photoz2} flag of 0 and another 59.0 million with a flag of 2.
In the validation set, 68\% of flag $=0$ galaxies have photometric redshift
within 0.026 of the measured spectroscopic redshift (in the range
$0.001 \leq z \leq 1.5$).  The rms dispersion between photometric and 
spectroscopic redshifts is higher, $\sigma=0.039$, a consequence of the 
non-Gaussian tails of the error distribution.

\begin{figure}
\epsscale{0.4}
\plotone{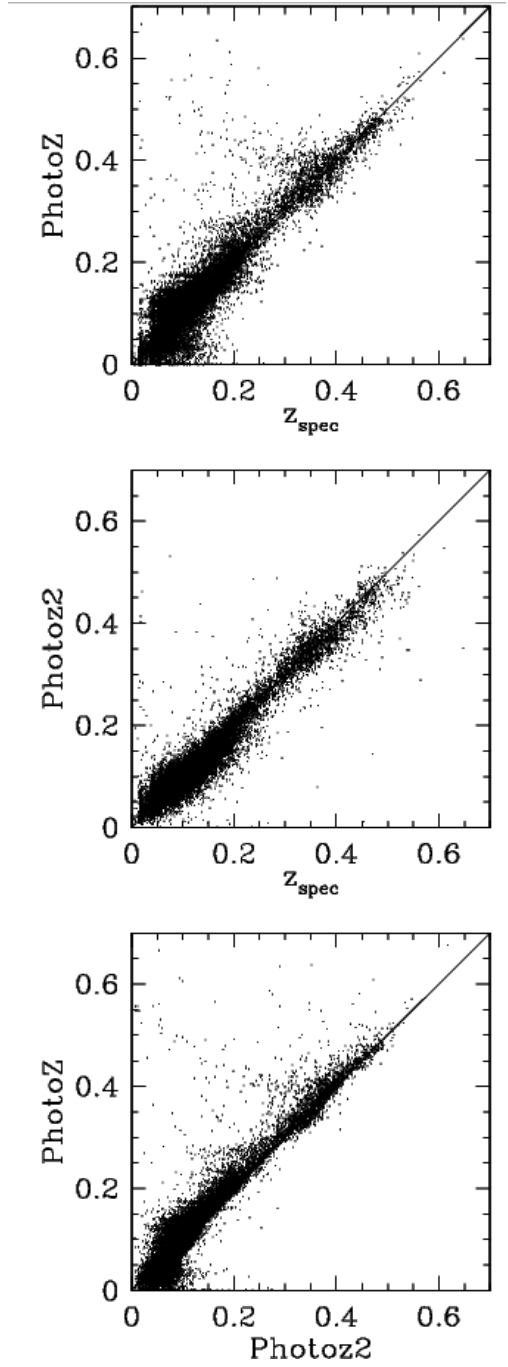}
\caption{
Comparison of photometric redshift estimates {\tt PhotoZ} and
{\tt PhotoZ2} to SDSS spectroscopic redshifts, and to each other.
\label{fig:photozcomp}}\end{figure}  

Figure~\ref{fig:photozcomp} plots the two photometric redshift
estimates against spectroscopic redshifts, and against each other,
for 20,000 objects selected from the DR5 database.  These are
objects with SDSS spectroscopic redshifts, spectroscopically 
classified as galaxies, {\tt PhotoZ} quality 
flag of 4 or 5, and {\tt PhotoZ2} flag of 0 or 2. 
Both estimates show a tight correlation with spectroscopic redshift
for the great majority of sources, while {\tt PhotoZ} shows a 
somewhat larger fraction of outliers with overestimated 
photometric redshifts.

\subsection{Regions and Sectors}
\label{sec:sector}

Each survey observation, imaging or spectroscopic, covers a 
certain region of the sky.  Doing statistical calculations with
the SDSS data usually requires performing computations over
these regions and the intersections among them,
e.g., to normalize luminosity functions or calculate
completeness corrections.
Typical questions are: how much area do these regions cover,
how much do they overlap, and which regions contain a
certain point or area of the sky?
The DR5 CAS includes tables that precisely describe each region
and built-in tools for finding
the connections and overlaps between one kind of region and another.
Each {\tt Region} in the CAS is represented
as a union of spherical polygons, and its area
is analytically calculated and stored.

The SDSS has many different types of regions; they include the stripes,
camera columns, segments, chunks, and spectroscopic tiles that
are the basis of the SDSS observing and target selection strategy.
The survey stripes overlap at the edges, with the overlap increasing
towards the survey poles, so they are clipped into disjoint
``staves'' centered on each stripe that uniquely cover the survey
area (like the staves of a barrel).  The union of the staves
within the survey boundaries defines the survey's ``primary''
photometric area.  There are ``holes'' inside the stripes and
staves, consisting of fields that were declared to be of
inferior quality (e.g., because of degraded seeing or 
contamination by the saturated pixels of a bright star and
its wings).  The portions of these holes that lie within
the primary survey area are called {\tt TiHoles} to emphasize
their role in the tiling process, as explained below.

As a simple example of the region tables, let us calculate
the photometric survey area.  Imaging data are imported to the database
in ``chunks,'' and the total area of these chunks can be obtained
from the SQL (Structured Query Language) query\footnote{See
{\tt http://cas.sdss.org/dr5/en/help/docs/sql\_help.asp}.
The text follows our standard capitalization conventions;
for example, the various types of entries in the {\tt Region}
table ({\tt CHUNK, TILE}, etc.) are
listed in all capital letters.  However, queries
are not case-sensitive.}

\smallskip
$\phantom{xxxxx}$ {\tt select sum(area) from Region where type=`CHUNK'},

\smallskip\noindent 
yielding 9560 deg$^2$.  However, this counts overlapping
areas more than once.  To obtain the unique survey imaging footprint,
we select only the ``primary'' region, the intersection of the
chunks with the staves,

\smallskip
$\phantom{xxxxx}$ {\tt select sum(area) from Region where type=`PRIMARY'},

\smallskip\noindent
yielding 7897 deg$^2$.  The total area
and unique footprint area should be adjusted downwards by the
area of the holes, obtained from the queries

\smallskip
$\phantom{xxxxx}$ {\tt select sum(area) from Region where type=`HOLE'}

\smallskip\noindent
for the chunks and 

\smallskip
$\phantom{xxxxx}$ {\tt select sum(area) from Region where type=`TIHOLE'}

\smallskip\noindent
for the primary area.  These queries yield 26 and 23 square
degrees, respectively, making the final precise numbers for the
photometric survey area 9534 deg$^2$ in total and 7875 unique deg$^2$
within the main survey boundaries.
(The 8000 deg$^2$ figure quoted elsewhere includes a small amount
of imaging outside of the ellipse that defines the main survey boundary.)

For analyses of spectroscopic samples, the issues are more complex.
The SDSS spectroscopic survey aims to sample quasars and galaxies
uniformly over the sky, with additional spectra for other samples
(not necessarily uniform) of science targets, calibration objects,
and sky.  In practice, after an area has been observed by the
photometric survey, a series of targeting pipelines creates lists
of targets that satisfy the selection criteria.  A ``tiling''
program (Blanton \etal\ 2003) runs over a subset of the observed
area and assigns targets to circular ``tiles'' of diameter
$2.98^\circ$; it also determines which targets are assigned fiber
holes on which spectroscopic plugplate, imposing physical 
constraints such as the $55''$ minimum spacing between fibers.
A given run of the tiling program operates on the union of a set
of ``rectangular'' (in spherical coordinates) {\tt TilingGeometry} areas.

For calculations of galaxy or quasar clustering,
one needs to compute the completeness of the
spectroscopic sample as a function of sky position.
The natural scale on which to do this is that of a
{\tt SECTOR}, a region that is covered by a unique set of 
{\tt Tile} overlaps (e.g., by a particular spectroscopic plate,
or by two or more plates that overlap).
These are regions over which the completeness should be nearly
uniform (see, e.g., Figure~1 of Percival \etal\ [2007] and earlier
discussions by Tegmark \etal\ [2004] and Blanton \etal\ [2005]).
The {\tt Target} table lists (in the column {\tt target.regionID})
the {\tt SECTOR} for every object selected
by the spectroscopic target selection algorithms, regardless
of whether or not that object has been spectroscopically observed.
To find the {\tt SECTOR} for an object in the main table of 
spectroscopically observed objects, {\tt SpecObj},
one must first identify the corresponding entry
in the {\tt Target} table.  For example, the following query

\smallskip\noindent
$\phantom{xxxxx}$ {\tt select top 10 s.specObjID, t.regionID} \\
$\phantom{xxxxx}$ {\tt from SpecObj s join Target t} \\
$\phantom{xxxxxxx}$ {\tt on s.targetID = t.targetID} \\

\smallskip\noindent
returns the spectroscopic ID numbers and the {\tt SECTOR} numbers of
the first ten objects encountered in the {\tt SpecObj} table.
The database function {\tt fRegionsContainingPointEQ}
can be used to find the {\tt SECTOR} that covers a specified point
on the sky.

The following practical example illustrates several other
features of these tables.
The SDSS quasar target selection algorithm underwent significant
changes in the early phases of the survey, reaching its final
form (Richards \etal\ 2002) with {\tt targetVersion} 3.1.0,
following DR1.  A calculation of the quasar luminosity function
should therefore be restricted to regions targeted with 
this or subsequent versions of the target selection code,
and it should be normalized using the corresponding area,
which the following query shows to be 4013 deg$^2$:

\smallskip\noindent
$\phantom{xxxxx}$ {\tt select sum(area)} \\
$\phantom{xxxxx}$ {\tt from Region} \\
$\phantom{xxxxx}$ {\tt where regionID in (} \\
$\phantom{xxxxxxxxxxx}$ {\tt select b.boxID} \\
$\phantom{xxxxxxxxxxx}$ {\tt from Region2Box b join TilingGeometry g} \\
$\phantom{xxxxxxxxxxxxxxxxx}$ {\tt on b.id = g.tilingGeometryID} \\
$\phantom{xxxxxxxxxxx}$ {\tt where b.boxType = `SECTOR'} \\
$\phantom{xxxxxxxxxxxxxxxxx}$ {\tt and b.regionType = `TIPRIMARY'} \\
$\phantom{xxxxxxxxxxx}$ {\tt group by b.boxID} \\
$\phantom{xxxxxxxxxxx}$ {\tt having min(g.targetVersion) >= 'v3\_1\_0' } \\
$\phantom{xxxxx}$ {\tt )} \\

\smallskip\noindent
This query uses the {\tt Region2Box} table, which maps between various
types of {\tt Regions} and the {\tt TilingGeometries} in which information
about the target selection is stored.  
The {\tt where} clause selects, from the table of all {\tt Regions},
those which are {\tt SECTOR}s in the primary tiled area and were
targeted with a final version of the quasar target selection 
algorithm.\footnote{This 
query is included as one of the sample queries in the DR5
documentation, under ``Uniform Quasar Sample,'' together with
a longer query that shows how to extract all quasars and
quasar candidates from the corresponding sky area.}

In principle, these tables provide all the information needed
for complex clustering calculations --- e.g., determining
local completeness corrections, generating appropriate
catalogs of randomly distributed points, and identifying
targeted objects that were not observed because of the
minimum fiber spacing constraint.  The queries required for
such calculations are rather lengthy, and will be 
presented and documented elsewhere.

\subsection{Match Tables}
\label{sec:match}

About 50 million photometric objects in the CAS lie in regions
that have been observed more than once, because of stripe overlap
or repeat scans.
These repeat observations can be used to detect variable and
moving objects.  The {\tt MatchHead} and {\tt Match} tables of
the DR5 CAS provide convenient tools to examine the multiple
observations of a single object, identified by positional matches
with a $1''$ tolerance, and collectively referred to as a {\it bundle}.
The {\tt MatchHead} table has the unique ID of the first object
in the bundle (defined by observation date), 
the mean and variance of the coordinates of all
matched detections, the number of matched detections, and the
number of times the object was ``missed'' in other observations
of the same sky area.  
Misses can occur because the object is variable, because it is
moving, because inferior seeing moves it below the detection threshold,
or because the original detection was spurious.
The {\tt Match} table lists all objects in each bundle.

As an example, the following query lists information about the 
multiple detections of an object at (ra,dec)=(194,0):

\smallskip
$\phantom{xxxxx}$ {\tt select MH.*} \\
$\phantom{xxxxxxxxxxx}$ {\tt from MatchHead MH} \\
$\phantom{xxxxxxxxxxx}$ {\tt join fGetNearbyObjEq(194,0,0.3) N on 
                             MH.objID = N.objID} \\
\smallskip

\noindent The {\tt fGetNearbyObjEq} function returns a table (assigned
the name {\tt N})
of all objects found within 0.3 arc-minutes of the desired coordinates.
The {\tt select} command returns all entries in the {\tt matchHead}
table (assigned the name {\tt MH}) which, as a result of the {\tt join}
command, have an object ID that matches one returned by the 
neighborhood search.  In this case, there is just one such match,
hence a single bundle.  One can get information on all the objects
in the bundle with the query

\smallskip
$\phantom{xxxxx}$ {\tt select M.*} \\
$\phantom{xxxxxxxxxxx}$ {\tt from Match M} \\
$\phantom{xxxxxxxxxxx}$ {\tt join MatchHead MH on M.matchHead = MH.objID} \\
$\phantom{xxxxxxxxxxx}$ {\tt join fGetNearbyObjEq(194,0,0.3) N on 
                             MH.objID = N.objID} \\
\smallskip

\noindent where the new {\tt join} command selects out those
{\tt Match} tables whose {\tt matchHead} agrees with that
returned by the earlier query.

The DR5 CAS has 50,627,023
bundles described by {\tt MatchHead} and
109,441,410 objects in the {\tt Match} table.
When an object is undetected in a repeat observation of
the same area of sky, a surrogate object is placed
in the {\tt Match} table but marked as a ``miss,'' with an additional flag
to indicate if the miss could be caused by masking of the
region in the second observation (e.g., because of a 
satellite trail or cosmic ray hit) or because it lies near
the edge of the overlap region.
A bundle may therefore consist of a single detection and one
or more surrogates (and the object in the {\tt MatchHead} may
be a surrogate).  
There are 9.8 million surrogates in the {\tt Match} table.  
The presence of surrogate objects 
may simplify algorithmic searches for moving or variable objects.

Because the multiple imaging scans of the southern equatorial stripe
are not yet in the CAS, the {\tt Match} tables cannot be used
to search for moving or variable objects in these data.
However, this capability will be present in future data releases.

\section{Conclusions}
\label{sec:conclusions}

The Fifth Data Release of the Sloan Digital Sky Survey provides access
to 8000 deg$^2$ of five-band imaging data and
over one million spectra.  These data
represent a roughly 20\% increment over
the previous data release (DR4, Adelman-McCarthy \etal\ 2006).  
Both the catalog data and the source imaging data are available
via the Internet.
All the data products have been consistently processed
by the same set of pipelines across several data releases.
The previous data releases remain online and unchanged to
support ongoing science studies.
DR5 includes several qualitatively new features: multiple imaging
scans of the southern equatorial stripe, special imaging scans of
M31 and the Perseus cluster, database access to
QSO catalogs and galaxy photometric redshifts, and database
tools for precisely defining the survey geometry and for
linking repeat imaging observations of matched objects.
More than a thousand scientific publications have been based
on the SDSS data to date, spanning an enormous range of subjects.
Future data releases will increase the survey area, and they
will provide
qualitatively new kinds of data on the stellar kinematics
and populations of the Milky Way and on Type Ia supernovae
and other transient or variable phenomena, further extending
this scientific impact.

\bigskip

We dedicate this paper to our colleague Jim Gray, who disappeared
in January, 2007, while sailing near San Francisco.
Jim dedicated an enormous amount of his time, his energy, and his
remarkable talents to the SDSS over the course of many years.
He played a critical role in the development of the SDSS database,
including important contributions to the writing of this paper.

Funding for the SDSS and SDSS-II has been provided by the Alfred P. Sloan 
Foundation, the Participating Institutions, the National Science Foundation, 
the U.S. Department of Energy, the National Aeronautics and Space 
Administration, the Japanese Monbukagakusho, the Max Planck Society, 
and the Higher Education Funding Council for England. The SDSS Web Site 
is http://www.sdss.org/.

The SDSS is managed by the Astrophysical Research Consortium for the 
Participating Institutions. The Participating Institutions are the 
American Museum of Natural History, Astrophysical Institute Potsdam, 
University of Basel, University of Cambridge, Case Western Reserve University, 
University of Chicago, Drexel University, Fermilab, the Institute for 
Advanced Study, the Japan Participation Group, Johns Hopkins University, 
the Joint Institute for Nuclear Astrophysics, the Kavli Institute for 
Particle Astrophysics and Cosmology, the Korean Scientist Group, 
the Chinese Academy of Sciences (LAMOST), Los Alamos National Laboratory, 
the Max-Planck-Institute for Astronomy (MPIA), the Max-Planck-Institute 
for Astrophysics (MPA), New Mexico State University, Ohio State University, 
University of Pittsburgh, University of Portsmouth, Princeton University, 
the United States Naval Observatory, and the University of Washington.

\end{document}